\newif\iffigs\figsfalse
\figstrue
\def\beq{\begin{equation}} 
\def\eeq{\end{equation}} 
\def\bed{\begin{displaymath}} 
\def\eed{\end{displaymath}} 
\def\beqq{\begin{eqnarray}} 
\def\eeqq{\end{eqnarray}} 
\def\bedd{\begin{eqnarray*}} 
\def\eedd{\end{eqnarray*}}

\def\n{\nonumber}
 
\def\bbbz{{\bf Z}} 
\def\bbbc{{\bf C}} 
\def\bbbr{{\rm I\!R}} 
 
\def\bbb1{{\rm 1\!1}}

\newcommand{\refs}[1]{(\ref{#1})}

\documentstyle[12pt]{article} 
 \input{epsf}
\advance\hoffset by -.35in 
\begin{document} 
\bibliographystyle{unsrt} 

\def\doublespace{\baselineskip=\normalbaselineskip \multiply\baselineskip by 2}
\def\singlespace{\baselineskip=\normalbaselineskip}


\def\pr{\prime} 
\def\pa{\partial} 
\def\es{\!=\!} 
\def\ha{{1\over 2}} 
\def\>{\rangle} 
\def\<{\langle} 
\def\mtx#1{\quad\hbox{{#1}}\quad} 
\def\pan{\par\noindent} 
\def\lam{\lambda} 
\def\La{\Lambda} 
 
\def\A{{\cal A}} 
\def\dal{\tilde\alpha}
\def\de{\delta}
\def\be{\beta}
\def\ga{\gamma}
\def\G{\Gamma} 
\def\Ga{\Gamma} 
\def\F{{\cal F}} 
\def\J{{\cal J}} 
\def\M{{\cal M}} 
\def\R{{\cal R}} 
\def\W{{\cal W}} 
\def\tr{\hbox{tr}} 
\def\al{\alpha} 
\def\d{\hbox{d}} 
\def\De{\Delta} 
\def\L{{\cal L}} 
\def\H{{\cal H}} 
\def\Tr{\hbox{Tr}} 
\def\I{\hbox{Im}} 
\def\e{\epsilon}
\def\R{\hbox{Re}} 
\def\h{\bar h}
\def\di{\partial\!\!\!\!\,/}
\def\ti{\int\d^2\theta} 
\def\bti{\int\d^2\bar\theta} 
\def\ttbi{\int\d^2\theta\d^2\bar\theta} 
 \def\bD{\bar D}
\def\bh{\bar h}
\def\la{\lambda}
\def\bp{\bar\phi}
\def\W{{\cal{W}}}
\def\La{\Lambda} 
\def\Ga{\Gamma} 
\def\ga{\gamma}
\def\dga{\tilde\gamma}
\def\be{\beta}
\def\dbe{\tilde\beta}
\def\W{{\cal W}} 
\def\L{{\cal L}}
\def\tr{\hbox{tr}} 
\def\al{\alpha} 
\def\dal{\tilde\alpha}
\def\De{\Delta} 
\def\de{\delta}
\def\L{{\cal L}} 
\def\nn{\nonumber}
\def\la{\lambda}
\def\cov{{\bf D}}
\def\Om{\Omega}
\def\Di{\cov\!\!\!\!/}
      

\begin{titlepage}
\hfill{DPT/99/73}
\vspace*{2cm}
\begin{center}
{\Large\bf Microscopic- versus Effective Coupling 
in N=2 Yang-Mills With Four Flavours}\\
\vspace*{2cm}
Ivo Sachs and William Weir\\
\vspace*{1cm}
{\em Department of Mathematical Sciences\\
University of Durham\\
Science Site, Durham DH1 3LE, UK}\\
\vspace*{2cm}
\end{center}
\begin{abstract} 
We determine the instanton corrections to the effective coupling in 
$SU(2)$, $N\es 2$ Yang-Mills theory with four flavours to all orders. 
Our analysis uses the $S(2,Z)$-invariant curve and the two instanton 
contribution obtained earlier to fix the higher order contributions 
uniquely. 
\end{abstract}
\end{titlepage}

\section{Introduction}
Seiberg and Witten \cite{SW1,SW2} proposed 
exact results for $SU(2)$, $N\es 2$ supersymmetric Yang-Mills theory with and 
without matter multiplets. These include, in particular an exact 
expression for the mass spectrum of BPS-states for these theories. Their solutions 
also provide a mechanism, based on monopole condensation, 
for chiral symmetry breaking and confinement in $N\es 2$ Yang-Mills with and 
without coupling to fundamental hypermultiplets 
respectively. The results of Seiberg and Witten have since been generalised to 
a variety of gauge groups 
\cite{su(n)} describing a number of interesting new phenomena \cite{new}. 

On another front the low energy effective theories arising 
from non-Abelian YM-theory have been identified with those describing the low 
energy dynamics of certain intersecting brane configurations in string 
theory \cite{branes}. This approach to field theory has the advantage of 
providing an elegant geometrical representation of the low energy dynamics of strongly coupled supersymmetric gauge theory. 

For gauge groups $SU(2)$ and $N_{F}\leq 3$ massless flavours it has 
since been shown that the solution \cite{SW1,SW2} are indeed the only ones 
compatible with supersymmetry and asymptotic freedom in these theories 
\cite{M1}. However, a number of issues have 
still resisted an exact treatment. In particular the 
precise relation between the low energy effective coupling 
$\tau_{\hbox{\tiny eff}}$  
and the microscopic coupling $\tau$ in the scale invariant $N_{F}\es 4$ 
theory has not been understood so far. Indeed, while the 
two couplings were first assumed to be identical in 
\cite{SW2} it was later found by explicit computation that there are, in 
fact, perturbative as well as 
instanton corrections \cite{Valya}. On the other hand, explicit instanton 
calculus is so far limited to topological charge $k\leq 2$. A related 
observation has been made in the D-brane approach to scale invariant 
theories. The details of the conclusions reached there are, however, 
somewhat different \cite{Randall}.

The purpose of the present paper is to fill this gap. Our analysis 
uses a combination of analytic results from the theory of conformal 
mappings combined with the known results form instanton calculus. 
More precisely we consider the sequence 
$\tau\mapsto z\in\bbbc\mapsto\tau_{\hbox{\tiny eff}}$. We will then argue that given 
the Seiberg-Witten curve together with some suitable assumptions on 
the singular behaviour of the instanton contributions there is a 
one-parameter family of admissible maps $\tau_{\hbox{\tiny eff}}\mapsto\tau$. The 
remaining free parameter is in turn determined by the two-instanton 
contribution to the asymptotic expansion at weak coupling. This 
coefficient has been computed explicitly in \cite{Valya}. 
Combining these results then determines the map completely. Although we are not 
able to give a closed form of the map $\tau\mapsto\tau_{\hbox{\tiny eff}}$ 
globally the higher order instanton coefficients can be determined 
iteratively. We further discuss some global properties of the map 
qualitatively. In particular we will see that it is not single valued 
meaning that the instanton corrections lead to a cut in the strong 
coupling regime. In this note we restrict ourselves to gauge group $SU(2)$ 
leaving the extension to higher groups \cite{higher} for future work. 

\section{Review of N=2 Yang-Mills with 4 Flavours}

To prepare the ground let us first review some of the relevant 
features of the theory of interest \cite{SW2}, that is 
$N\es2$ YM-theory with $4$ hypermultiplets  $Q^{r}$ and $\tilde Q_{r}$, 
$r\es 1,\cdots,4$, in the 
fundamental representation. In $N\es 1$ language the hypermultiplets 
are described by two chiral multiplets containing the left handed quarks and 
anti quarks respectively. These are in isomorphic 
representations of the gauge group $SU(2)$. The global symmetry group 
therefore contains a $SO(8)$ or, more precisely, a $O(8)$ due to 
invariance under the $\bbbz_{2}$ ``parity'' which exchanges a left handed quark 
with its anti particle, $Q_{1}\leftrightarrow\tilde Q_{1}$, with all other fields 
invariant. At the quantum level this $\bbbz_{2}$ is anomalous due to 
contributions from odd instantons. 

We consider the Coulomb branch with a constant scalar $\varphi\es a\neq 0$ 
in the $N\es 2$ vector multiplet. This breaks the gauge group 
$SU(2)\to U(1)$. The charged hypermultiplets 
then have mass $M\es\sqrt{2}|a|$ and transform as a vector under 
$SO(8)$, rather than $O(8)$ due to the $\bbbz_{2}$-anomaly. 
In addition, there are magnetic monopole solutions 
leading to $8$ fermionic zero modes from the $4$ hypermultiplets. 
These turn the monopoles into spinors of $SO(8)$. The symmetry 
group is therefore the universal cover of $SO(8)$ or Spin$(8)$ 
with centre $\bbbz_{2}\times\bbbz_{2}$. Following \cite{SW2} we 
label the representations $\bbbz_{2}\times\bbbz_{2}$ according to 
the Spin$(8)$ representations, that is the trivial representation 
$o$, the vector representation $v$ and the two spinor representations  
$s$ and $c$. To decide in which spinor representation the monopoles 
and dyons transform one considers the action of an electric charge 
rotation $e^{\pi i Q}$ on these states. Here the electric charge $Q$ is 
normalised such that the massive gauge bosons have charge $\pm 2$. This action 
is conveniently described by 
\beq
e^{\pi i Q}=e^{{in_{m}\theta}}(-1)^{H},
\eeq
where states with even, odd $n_{e}$ are $(-1)^{H}$ even, odd 
respectively. On the other hand, for consistency, the 
monopole anti-monopole annihilation process requires a correlation 
between chirality in Spin$(8)$ and electric charge \cite{SW2}. We 
therefore identify $(-1)^{H}$ with the chirality operator in the spinor 
representations of Spin$(8)$. Hence dyons with even and odd electric charge 
transform in one or the other spinor representation of Spin$(8)$ respectively. 
There is an outer automorphism $S_{3}$ 
of Spin$(8)$ that permutes the three non-trivial representations 
$v$, $s$ and $c$. It is closely connected to the proposed duality group 
$SL(2,\bbbz)$ of the quantum theory. Indeed there is a homomorphism 
$SL(2,\bbbz)\to S_{3}$, so that the invariance group of the spectrum 
is given by the semi direct product Spin$(8)$ and $SL(2,\bbbz)$. 
The kernel of this homomorphism plays an important part in our analysis 
below. It consists of the matrices congruent to $1$ (mod$(2)$). They are conjugate to 
the subgroup $\Gamma_{0}(2)$. The fundamental domain of this 
subgroup is the space of inequivalent effective couplings $\tau_{\hbox{\tiny eff}}$. 

One can further formalise this structure in terms of the 
hyperelliptic curve that controls the low energy behaviour 
of the model \cite{SW2}. For this one seeks a curve $y^{2}\es 
F(x,u,\tau)$ such that the differential form 
\beq
\omega=\frac{\sqrt{2}}{8\pi}\frac{\d x}{y}
\eeq
has the periods 
$(\frac{\pa a_{D}}{\pa u},\frac{\pa a}{\pa u})$ with $(a_{D},a)$ 
given by \cite{SW2}
\beq
a=\sqrt{\frac{u}{2}}\mtx{and} a_{D}=\tau_{\hbox{\tiny eff}} a.
\eeq
The curve consistent with $SL(2,\bbbz)$ duality determined in \cite{SW2}, 
is given by 
\beq
y^{2}=(x-ue_{1}(\tau))(x-ue_{2}(\tau))(x-ue_{2}(\tau)),
\eeq
where $e_{i}(\tau)$ are the modular forms corresponding to the 
three subgroups of $SL(2,\bbbz)$, conjugate to the index $3$ subgroup 
$\Gamma_{0}(2)$. 

\section{Map: $\tau\mapsto\tau_{\hbox{\tiny eff}}$}

We now have the necessary ingredients to 
determine the precise relation between $\tau_{\hbox{\tiny eff}}$ and $\tau$. We 
begin with the observation that, according to the structure of the 
effective theory presented above, the fundamental domain $D_{\Gamma}$ 
of any of the three subgroups conjugate to $\Gamma_{0}(2)$ can be used 
as the space of inequivalent effective couplings. The three choices 
are then related by the Spin$(8)$ ``triality'' relating the  $3$ different 
non-trivial representations $v$, $s$ and $c$. Each fundamental domain 
is described by a triangle in the upper half plane ($\I(\tau)\geq 
0$), bounded by circular arcs \cite{Nehari}. The $3$ singularities are conjugate to the 
points $(i\infty,-1,1)$ 
corresponding to the weak coupling regime, massless monopoles and 
massless dyons with charge $(n_{m},n_{e})\es(1,1)$ mod$(2)$ 
respectively. 

In the absence of perturbative- and instanton corrections the 
effective coupling is identified with the microscopic coupling 
$\tau$. This is the case in $N\es 4$ theories. In the scale invariant $N\es 2$ 
theory considered here the situation is different. As shown in 
\cite{Valya}, the effective coupling is finitely renormalised at the 
one loop level and furthermore receives instanton corrections. As a result 
the fundamental domain $\bar D$, of inequivalent microscopic couplings and 
the fundamental domain $D_\Gamma$, of inequivalent effective couplings 
are not the same. We will now determine the exact relation between them. 

Some information about the fundamental domain $\bar D$ of microscopic 
couplings $\tau$ is obtained from the following 
observations:\hfill\break
{\it a)}\hspace{.2cm} As the microscopic coupling does not enter in the mass formula, its 
fundamental domain is not constrained to be that of a subgroup of 
$SL(2,\bbbz)$ \cite{M1}. Nevertheless we require that the imaginary 
part of $\tau$ be bounded from below. Correspondingly the different 
determinations of $\tau$ for a given $\tau_{\hbox{\tiny eff}}$ must be related by a 
transformation in $G\subset PSL(2,\bbbr)$. Hence, a particular determination of $\tau$. 
will lie within a fundamental domain of $G$ or some covering thereof. That is, 
\beq
\tau\in C[H/G]\mtx{where} G\subset PSL(2,\bbbr),
\eeq
where $C[\;]$ denotes a certain covering. Any such domain is bounded by 
circular arcs and is thus conformally equivalent to the punctured $2$-sphere, 
$\bbbc-\{a_{1},\cdots,a_{n}\}$ \cite{Nehari}.\hfill\break 
{\it b)}\hspace{.2cm} We know of no principle excluding the possibility that the number of 
vertices of $\bar D$ be different from that of $D_{\Gamma}$. On 
the other hand such extra singularities have no obvious physical 
interpretation. We therefore discard this possibility. 

Equipped with this information we will now determine the homomorphism that 
maps $D_{\Gamma}$ into the fundamental domain of microscopic 
couplings\footnote{As will become clear below the two domains cannot be 
isomorphic.} $\bar D$. It follows from general arguments 
\cite{SW2,Valya} that this map has an expansion of the form 
\beq\label{tautaue}
\tau_{\hbox{\tiny eff}}=\sum\limits_{n=0}^{\infty}c_{n}q^{n}\mtx{where} q=e^{\pi 
i\tau}
\eeq
The coefficients $c_{i}$, represent the perturbative one-loop  
($i\es 0$) and instanton ($i\!>\!0$)  corrections respectively. The 
contributions from odd instantons to $\tau_{\hbox{\tiny eff}}$ vanishes. This is 
due the fact that the part of the effective action determining the 
effective coupling is invariant under the $\bbbz_{2}$-''parity'' 
described in the last section. The 
first two non-vanishing coefficients of the expansion \refs{tautaue} are known 
\cite{Valya}. They are
\beq
c_0 = \frac{i}{\pi} 4 \ln 2 \mtx{and} c_2= -\frac{i}{\pi}  \frac{7}{2^5
\cdot 3^6}.
\eeq

To continue we use some elements of the theory of conformal mappings 
\cite{Nehari}. That is we consider the maps from the punctured 
$2$-sphere $S=\bbbc-\{a_{1},\cdots,a_{n}\}$ to polygons in the upper half plane, 
bounded by circular arcs. Concretely we consider the 
sequence $\tau_{\hbox{\tiny eff}}\mapsto z\in S\mapsto\tau$ (see Fig.1). The form of such  
mappings is generally complicated. However, their 
Schwarzian derivative takes a remarkably simple form \cite{Nehari}
\beq\label{sw0}
\{\tau,z\}=\sum\limits_{i}\ha\frac{1-\al_{i}^{2}}{(z-a_{i})^{2}}
+\frac{\be_{i}}{z-a_{i}}.
\eeq
The parameters $\al_{i}$ measure the angles of the polygon in units 
of $\pi$. The accessory parameters $\be_{i}$ do not have a simple 
geometric interpretation but are determined uniquely 
up to a $SL(2,\bbbc)$ transformation of $S$. Furthermore they 
satisfy the conditions \cite{Nehari}
\beqq\label{sw1}
\sum_{i=1}^{n} \be_{i}  = 0&,&\sum_{i=1}^{n} \left[ 2a_{i} \be_{i} + 1 - \al_{i}^{2} \right]  =  0 ,\n\\
\sum_{i=1}^{n} \left[ \be_{i}a_{i}^{2}+a_i \left( 1 - \al_{i}^{2} \right) 
\right] & = & 0 
\eeqq
In the present situation it is convenient to orient the polygons 
such that they have a vertex at infinity with zero angle (see Fig. 1 ) 
corresponding to the weak coupling singularity 
$(\tau\es\tau_{\hbox{\tiny eff}}\es i\infty)$. 
The above conditions then simplify to 
\beqq\label{cs}
\sum\limits_{i=1}^{\infty}\be_{i}&=& 0\\еее
\sum_{i=1}^{n-1} \left(2 a_i \be_i -\al_{i}^{2} \right) &=& (2-n).\n
\eeqq

As explained at the beginning of this section, in the case at hand, 
the polygon on the $\tau_{\hbox{\tiny eff}}$-side corresponds to the fundamental 
domain of $\Gamma_{0}(2)$. The corresponding parameters 
$(a_{i},\al_{i},\be_{i})$ are given by \cite{M1}
\beq
a_{1}=a_{-1}=-1,\qquad \be_{1}=\be_{-1}=-\frac{1}{4},\qquad \al_{i}=0.
\eeq
The parameters for the polygon on the $\tau$-side, $(\tilde a_{i},\tilde\al_{i},
\tilde\be_{i})$ are to be determined.
However, the conditions \refs{cs} together with the symmetry 
$\tau\to-\bar\tau$ leaves only one free parameter. Indeed, without 
restricting the generality we 
can choose $\tilde a_{i}\es a_{i}$. Furthermore 
$\tilde\al_{i}\es-\tilde \al_{i}$. Then \refs{cs} implies 
\beq
\tilde\be_{-1} =  -\tilde\be_1 = \frac{1}{4} \left( 1-2\tilde\al_{1}е^2 \right)
\eeq
leaving only one parameter, $\tilde\al_{1}$ say, undetermined. As we shall 
now see this parameter is in turn determined by the two instanton 
contribution in \refs{tautaue}. For this we make use of the 
identity 
\beqq\label{tt}
\{\tau,\tau_{\hbox{\tiny eff}}\}\left(\frac{\pa\tau_{\hbox{\tiny eff}}}{\pa z}\right)^{2}&=&\{\tau,z\}-
\{\tau_{\hbox{\tiny eff}},z\}\\
&=&\sum_{i=1}^{{\tilde n}-1} \ha
\frac{1-{\tilde \al}_{i}^{2}}{(z-{\tilde a}_i)^2} +
\frac{{\tilde \be}_{i}}{(z-{\tilde a}_i)}
-\sum_{i=1}^{n-1} \ha
\frac{1-\al_{i}^{2}}{(z-a_i)^2} + \frac{\be_{i}}{(z-a_i)}.\n
\eeqq
To continue we invert \refs{tautaue} as 
\beq
\tau = \tau_{\hbox{\tiny eff}} - c_0 -c_2 e^{-2 \pi i c_0} q_{\hbox{\tiny eff}} + \left(2 \pi i 
c_{2}^{2}
- c_4 \right) e^{-4 \pi i c_0} q_{\hbox{\tiny eff}}^2 + O(q_{\hbox{\tiny eff}}^{4}).
\eeq
Finally we need the form of $\tau_{\hbox{\tiny eff}}(z)$, that is the inverse 
modular function for $\Gamma_{0}(2)$ \cite{Erdely,M1}
\begin{equation}
\label{tez}
\tau_{\hbox{\tiny eff}} = i \frac{ _2 F_1 \left( \ha , \ha ; 1 ;
\frac{-1+z}{1+ z} \right)  }{ _2 F_1 \left( \ha, \ha ; 1;
\frac{2}{1+ z} \right) }
\end{equation}
This function has the asymptotic expansion for large $z$
\begin{equation}
\label{tas}
\tau_{\hbox{\tiny eff}}(z)  = \frac{i}{\pi} \left[ 3 \ln 2 + \ln z - \frac{5}{16} \frac{1}{z^2}
\right] + O(z^{-3}).
\end{equation}
Substituting \refs{tas} into the right hand side of \refs{tt} we end 
up with 
\begin{equation}
\label{ttel}
- \left\{ \tau , \tau_{\hbox{\tiny eff}} \right\} \left( \frac{\pa\tau_{\hbox{\tiny eff}}}{\pa z} \right)^2 =
\frac{7}{2 \cdot 3^5} \frac{1}{z^4} + \left( \frac{7 \cdot 7213}{2^5 \cdot
3^{10} } +
2^{10} i \pi c_4 \right) \frac{1}{z^6}
\end{equation}
Substitution of \refs{ttel} into \refs{tt} then leads to 
\begin{equation}
\label{tal}
{\tilde \al_{1}е}^2 = \frac{7}{2^2 \cdot 3^5}
\end{equation}
which then fixes the erstwhile free parameter in $\left\{ \tau , z
\right\}$. This is the result we have been aiming at. Indeed all higher 
instanton coefficients are now determined implicitly by the 
equation 
\beq\label{tt2}
\{\tau,\tau_{\hbox{\tiny eff}}\}=\left(\frac{\pa z}{\pa\tau_{\hbox{\tiny eff}}}\right)^{2}\left[\{\tau,z\}-
\{\tau_{\hbox{\tiny eff}},z\}\right].
\eeq

\iffigs

\begin{figure}[t]
\hfill
\centerline{\epsffile{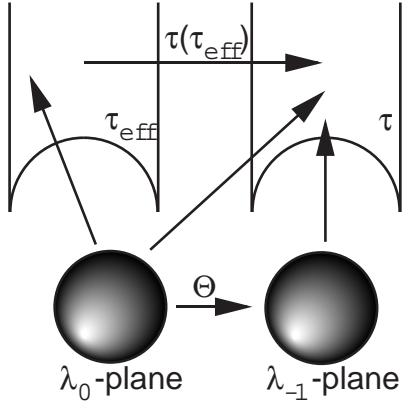}}
\caption{ Relation between $\tau_{\hbox{\tiny eff}}$ and $\tau$. $z=\lambda_{-1} =-1+2/\lambda_{0}$}
\hfill
\end{figure}

\else
\message{No figures will be included. See TeX file for more
information.}
\fi

In order to integrate \refs{tt} one  notices \cite{Erdely} that 
any solution of \refs{sw0} can be written as a quotient 
\beq\label{sol}
\tau(z)=\frac{(u_1(z)+du_2(z))}{(eu_1(z)+fu_2(z))},
\eeq
where $u_1,u_2$ are two linearly independent solutions of the hypergeometric differential equation 
\begin{equation}
\label{hyper1}
(1+z)(1-z) \frac{d^2}{dz^2} u(z) + \left( (c-2)z + c -2a -2b \right)
\frac{d}{dz}
u(z) + \frac{2ab}{1+z} = 0,
\end{equation}
with 
\beq
c=1,\mtx{;}b(c-a)+a(c-b)=\ha\mtx{and} (a-b)^2=\tilde\al_1^2.
\eeq
The coefficients $d,e,f$ in\refs{sol} are determined by the asymptotic 
expansion 
\beq
\tau(z)=\frac{i}{\pi}\ln\frac{z}{2}+\frac{i}{8\pi}z^{-2}\left(-\frac{5}{2}+\frac{7}{3^6}\right)+O(z^{-4}). 
\eeq
Finally we substitute $z$ in \refs{sol} by 
\beq
z(\tau_{\hbox{\tiny eff}})\equiv\lambda_{-1}= \frac{2}{\lambda_0(\tau_{\hbox{\tiny eff}})}-1,
\eeq
where $\lambda_0$ is the automorphic function 
\beqq
\tau_{\hbox{\tiny eff}}^{-1}&:& \tau_{\hbox{\tiny eff}}\mapsto \lambda_0\in\bbbc,\\
\lambda_0(\tau_{\hbox{\tiny eff}})&=&16 q_{e} \prod_{n=1}^{\infty} \left( \frac{1+q_{e}^{2n}}{1+q_{e}^{2n-1}}
\right)^8 \mtx{ with } q_{e}=\exp(i \pi \tau_{\hbox{\tiny eff}}).\n
\eeqq
This then integrates \refs{tt}. To extract the instanton coefficients one 
needs the inverse map $\tau_{\hbox{\tiny eff}}(\tau)$. This can be done iteratively. 
We have done this to $O(q^4)$ allowing us to predict the $4$-instanton 
coefficient 
\beq
c_4 = \frac{i}{\pi} \frac{ 7 \cdot 17 \cdot 421}{2^6 \cdot 3^{10} \cdot 521}.
\eeq

We close with the observation that globally the inverse function 
$\tau_{\hbox{\tiny eff}}(\tau)$ cannot be single valued. Indeed, existence of 
a single valued inverse function $z(\tau)$ requires $\tilde\al_1\es p$ or 
$\al\es 1/p$, $p\in \bbbz$ \cite{Erdely}. 
As a consequence, the instanton corrected effective coupling 
$\tau_{\hbox{\tiny eff}}(\tau)$ has a cut somewhere in the strong coupling region. 
It would certainly be interesting to understand the origin of this branch 
cut from the non-perturbative physics of this model. Thus far this remains 
elusive to us.

\vspace{1cm}

\pan
{\bf Acknowledgements:}\pan
\pan
I.S. would like to thank the Department of Mathematics at 
Kings College London for hospitality during the writing up of this 
work. I.S. was supported by a Swiss Government TMR Grant, BBW Nr. 
970557.

\end{document}

\end